\documentclass[a4paper,11pt]{article}

\usepackage{jheppub} 

\usepackage[T1]{fontenc}
\newcommand{\be}{\begin{equation}}
\newcommand{\ee}{\end{equation}}

\newcommand{\bea}{\begin{eqnarray}}
\newcommand{\eea}{\end{eqnarray}}
\newcommand{\bfig}{\begin{figure}}
\newcommand{\efig}{\end{figure}}
\newcommand{\bc}{\begin{center}}
\newcommand{\ec}{\end{center}}

\newcommand\scalemath[2]{\scalebox{#1}{\mbox{\ensuremath{\displaystyle #2}}}}
\newcommand{\ie}{i.e.\ }

\newcommand{\minus}{\ensuremath \scalebox{0.5}[1.0]{\( - \)}}
\newcommand{\pminus}{\hphantom{\minus}}

\DeclareFontFamily{U}{wncy}{}
\DeclareFontShape{U}{wncy}{m}{n}{<->wncyr10}{}
\DeclareSymbolFont{mcy}{U}{wncy}{m}{n}
\DeclareMathSymbol{\sha}{\mathord}{mcy}{"58}

\title{Two-Loop master integrals for heavy-to-light form factors of two different massive fermions}

\author[a]{Long-Bin Chen}

\affiliation[a]{School of Physics \& Electronic Engineering, Guangzhou University,
\\ 230 GuangZhou University City Outer Ring Road, Guangzhou 510006, China}

\emailAdd{chenglogbin10@mails.ucas.ac.cn}

\abstract{We calculate the full set of the two-loop master integrals for heavy-to-light form factors of two different massive fermions for arbitrary momentum transfer in NNLO QCD or QED corrections. These integrals allow to determine the two-loop QCD or QED corrections to the amplitudes for heavy-to-light form factors of two massive fermions in a full analytical way, without any approximations. The analytical results of the master integrals are derived using the method of differential equations, along with a proper choosing of canonical basis for the master integrals. All the results of master integrals are expressed in terms of Goncharov polylogarithms.}

\preprint{~}

\begin{document}
\maketitle
\flushbottom

\section{Introduction}

The calculation of heavy-to-light form factors  have a number of applications in particle physics. For instance, the semileptonic decay of a heavy quark to a light massive quark $(t\rightarrow b + W^+(l +\bar{\nu}) , b \rightarrow c +l +\bar{\nu})$, the decay of a massive lepton to anther massive lepton $(\mu \rightarrow e +\nu_{\mu} + \bar{\nu}_e,\tau \rightarrow \mu+\nu_{\tau} + \bar{\nu}_\mu)$ , and the decay of $W$ bosons into two massive quarks.  The NNLO QCD or QED corrections to the decay of a heavy fermion to a light fermion were computed in a number of papers \cite{Chetyrkin:1999ju,Blokland:2004ye,Czarnecki:2005vr,Brucherseifer:2013iv,Gao:2012ja,Bonciani:2008wf,Asatrian:2008uk,Beneke:2008ei,Bell:2008ws}, and the masses of the light fermions were neglected to simplify the calculation. In Ref. \cite{Pak:2008qt,Pak:2008cp,Dowling:2008ap}, the semileptonic decay $b \rightarrow c +l +\bar{\nu}$ have been calculated with the mass of charm quark taken into account, an expansion in powers and logarithms of mass ratio $\frac{m_c}{m_b}$ was performed there.

It has been shown that in the calculation of similar NNLO corrections to light fermion energy spectrum in heavy fermion decay, the mass of the light fermion can not be neglected since there exist the large logarithm terms $\ln(\frac{m_{\text{heavy}}}{m_{\text{light}}})$ \cite{Arbuzov:2002cn}. These large logarithm terms cancel out in the calculation of the total rate which make the calculation simpler. At order ${\cal O}(\alpha^{2})$ or ${\cal O}(\alpha_s^{2})$, double-logarithmic $\ln^2(\frac{m_{\text{heavy}}}{m_{\text{light}}})$ and single-logarithmic $\ln(\frac{m_{\text{heavy}}}{m_{\text{light}}})$ enhanced terms will appear, which make it impossible to compute the radiative corrections to quantities such as the light fermion energy spectrum by neglecting the mass of the light fermion from the very beginning. Having the masses as regulators can simplify the treatment of real emission processes, however the computation of virtual corrections is more complicate compared to a purely massless case. The full dependence on the muon and electron masses for electron energy spectrum in muon decay at NNLO QED corrections have first been calculated in a numerical way in \cite{Anastasiou:2005pn}. However, the appearance of $\frac{m^2_{\text{light}}}{m^2_{\text{heavy}}}\ll1$ will cause numerical instability for the numerical evaluation of multi-dimension integrals, which make the analytical calculations highly desirable. Moreover, the analytic expressions are obviously desirable in order to have control over errors in approximations, especially when the convergence of expansion in powers of mass ratio works slowly.

On the theoretical side, unravelling the mathematical structure of Feynman integrals will be important to handle the complexity of their calculation and may help us to obtain a better understanding of the perturbative quantum field theory. The study of the mathematical properties of Feynman integrals has attracted increasing attention both by the physics and the mathematics communities. Significant progresses were achieved in understanding the analytical computations of multi-loop Feynman integrals in the last years .

One of the powerful methods to evaluate the master integrals analytically is the method of differential equations \cite{Kotikov:1990kg, Kotikov:1991pm, Remiddi:1997ny, Gehrmann:1999as, Argeri:2007up}. Along with the recent years' development \cite{Henn:2013pwa,Henn:2013nsa,Henn:2014qga,Argeri:2014qva,Liu:2017jxz}, this method is becoming more and more powerful. It is pointed out in Ref. \cite{Henn:2013pwa} that for multi-loop Feynman integrals calculations, a suitable basis of master integrals can be chosen, so that the corresponding differential equations are greatly simplified, and their iterative solutions become straightforward in terms of dimensional regularization parameter $\epsilon=\frac{4-D}{2}$. The choice of canonical basis will also simplify the determination of boundary conditions considerably. Following this proposal, substantive analytical computations of various phenomenology processes have been completed \cite{Henn:2013woa,Henn:2014lfa,Gehrmann:2014bfa,Caola:2014lpa,DiVita:2014pza,Bell:2014zya,Huber:2015bva,Bonciani:2015eua,Gehrmann:2015dua,Grozin:2015kna,Bonciani:2016ypc,Mastrolia:2017pfy}.

The two-loop master integrals of  QED electron form factors have been calculated in \cite{Bonciani:2003te}, for on shell electrons of finite squared mass and arbitrary momentum transfer. The calculations of those master integrals have been refined by a suitable choice of basis \cite{Argeri:2014qva}. The analytical results of two-loop master integrals for form factors of heavy fermion to massless fermion have been obtained in \cite{Bonciani:2008wf,Bell:2006tz,Bell:2007tv,Huber:2009se}. All the master integrals for these two processes can be expressed in terms of Harmonic polylogarithms.  However, for two-loop vertex integrals with two different type of massive fermions, the master integrals will contain one more scale, to the best of our knowledge, the integrals have not been calculated analytically in the literature. Furthermore, understanding the structure of loop integrals more generally is an interesting and important challenge. In this work, employing the method of differential equations, along with a proper choice of canonical basis, we calculate all the master integrals for two-loop heavy-to-light form factors of two massive fermions, the results are expressed in terms of Goncharov polylogarithms.

The outline of the paper is as follows. In section 2, the kinematics and notations are introduced for the processes we concern. We also present the generic form of the differential equations with respect to the kinematics variables in terms of the derivatives of the external momentum. In section 3, the Goncharov polylogarithms as well as Harmonic polylogarithms are introduced. In section 4, the canonical basis are explicitly presented, followed by the discussion of their solutions. In sections 5, the determination of the boundary conditions, as well as the analytical continuation are explained. Discussions and conclusions are made in section 6. In Appendix A, we present all the rational matrices of the system of differential equations in canonical form. All the analytical results up to weight four from our computation as well as the rational matrices in electronic form are collected in ancillary files accompanying the \textbf{arXiv} version of this publication.

\section{Notations and Kinematics}

We consider the process of a heavy fermion decay into a light massive fermion, or the decay of W boson into two massive fermions,
\bea
f^1(p_1)\rightarrow V^*(q)+f^2(p_2),\label{pro1}\\
V(q) \rightarrow f^1(p_1)+f^2(p_2),\label{pro2}
\eea
with $p_1^2=m_1^2$ and $p_2^2=m_2^2$. For the decay of a heavy fermion to a light massive fermion, the squared momentum transfer have the following relations
\bea
q^2=(p_1-p_2)^2=s<(m_1-m_2)^2.
\eea
While for the decay of W boson into two different massive fermions, we have
\bea
q^2=(p_1+p_2)^2=s>(m_1+m_2)^2.
\eea

We take the dimensionless variables $x$ and $y$ to express the analytical results, they are defined by
\bea
s=m_1^2\frac{(x-y)(1-x\, y)}{x} ,\, \, \text{and } \,  m_2= m_1\, y.
\eea
In the above equations $m_1$ is treated as constants, $s$ and $m_2$ are considered as variable. The derivatives of $s$ and $m_2^2$ can be written in terms of
the derivatives of the external momenta and expressed as
\bea
\frac{\partial}{\partial s} &=& \frac{1}{(s-(m_1+m_2)^2)(s-(m_1-m_2)^2)}((s-m_1^2-m_2^2)p_1-2m_2^2\, p_2)\cdot\frac{\partial}{\partial p_1},\nonumber\\
\frac{\partial}{\partial m_2^2} &=& \frac{1}{(s-(m_1+m_2)^2)(s-(m_1-m_2)^2)}((-s-m_1^2+m_2^2)p_1+(s-m_1^2+m_2^2)p_2)\cdot\frac{\partial}{\partial p_1}.\nonumber\\
\eea

The corrections to the processes (\ref{pro1}) and (\ref{pro2}) could be calculated using Feynman diagram approach. All the amplitudes can be expressed in terms of a set of 40 scalar integrals. The calculation of these scalar integrals always turns out to be the toughest parts in the whole work. We use packages $\textbf{FIRE}$ \cite{Smirnov:2008iw,Smirnov:2013dia,Smirnov:2014hma} to reduce the group of scalar integrals into a minimum set of independent master integrals. $\textbf{FIRE}$ is also adopted in the derivations of differential equations for master integrals.

\section{Goncharov polylogarithms and Harmonic polylogarithms}

The Goncharov polylogarithms (GPLs) \cite{Goncharov:1998kja} we use to express the analytical results are defined as follow
\bea
G_{a_1,a_2,\ldots,a_n}(x) &\equiv & \int_0^x \frac{\text{d} t}{t - a_1} G_{a_2,\ldots,a_n}(x)\, ,\\
G_{\overrightarrow{0}_n}(x) & \equiv & \frac{1}{n!}\log^n x\, .
\eea

They can be viewed as a special case belonging to a more general type of functions named Chen-iterated integrals \cite{Chen}. If all the index $a_i$ belong to the set $\{0, \pm 1\}$, the Goncharov polylogarithms turn into the well-known Harmonic polylogarithms (HPLs) \cite{Remiddi:1999ew}
\bea
H_{\overrightarrow{0}_n}(x) &=&G_{\overrightarrow{0}_n}(x)\, ,\\
H_{a_1,a_2,\ldots,a_n}(x) &=&(-1)^k G_{a_1,a_2,\ldots,a_n}(x),
\eea
where $k$ equals to the times of element $(+1)$ taken in $(a_1,a_2,\ldots,a_n)$\, .

The Goncharov polylogarithms fulfil the following shuffle rules
\bea
G_{a_1,\ldots,a_m}(x)G_{b_1,\ldots,b_n}(x) &=& \sum_{c\in a \sha b} G_{c_1, c_2,\ldots,c_{m+n}}(x)\, .
\eea
Here, $a \sha b$ is composed of the shuffle products of lists a and b. It is defined as the set of
the lists containing all the elements of a and b, with the ordering of the elements
of a and b preserved. Both the GPLs and HPLs can be numerically evaluated within the {\bf GINAC} implementation \cite{Vollinga:2004sn,Bauer:2000cp}. A Mathematica package {\bf HPL} \cite{Maitre:2005uu,Maitre:2007kp} is also available to reduce and evaluate the HPLs.  Both the GPLs and HPLs  can be transformed to the functions of  $\ln, \text{Li}_n$ and $\text{Li}_{22}$  up to weight four, with the algorithms and packages described in \cite{Frellesvig:2016ske}.

\section{The canonical basis}

\begin{figure}[h]
\begin{center}
\includegraphics[scale=0.4]{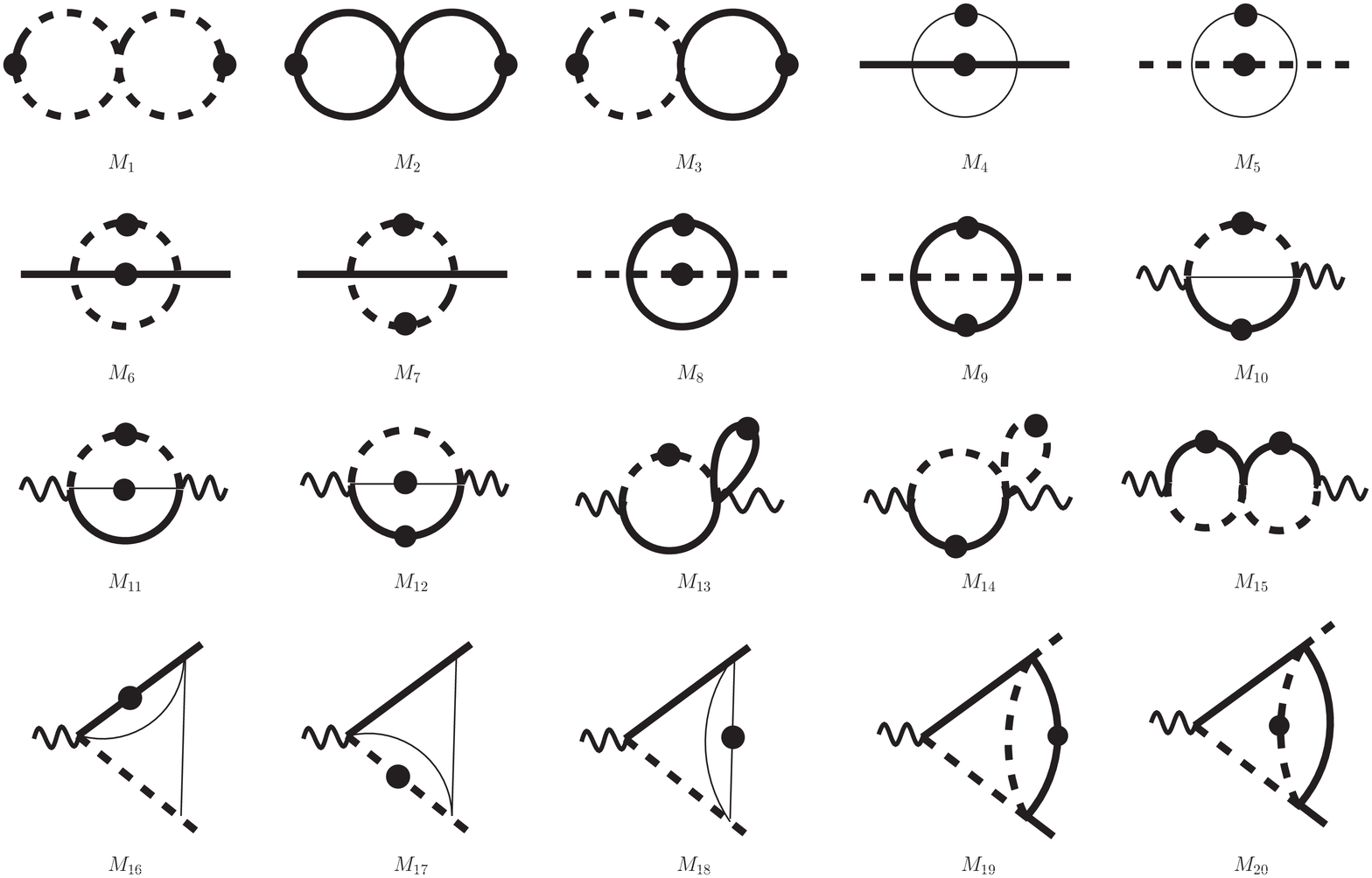}
\includegraphics[scale=0.4]{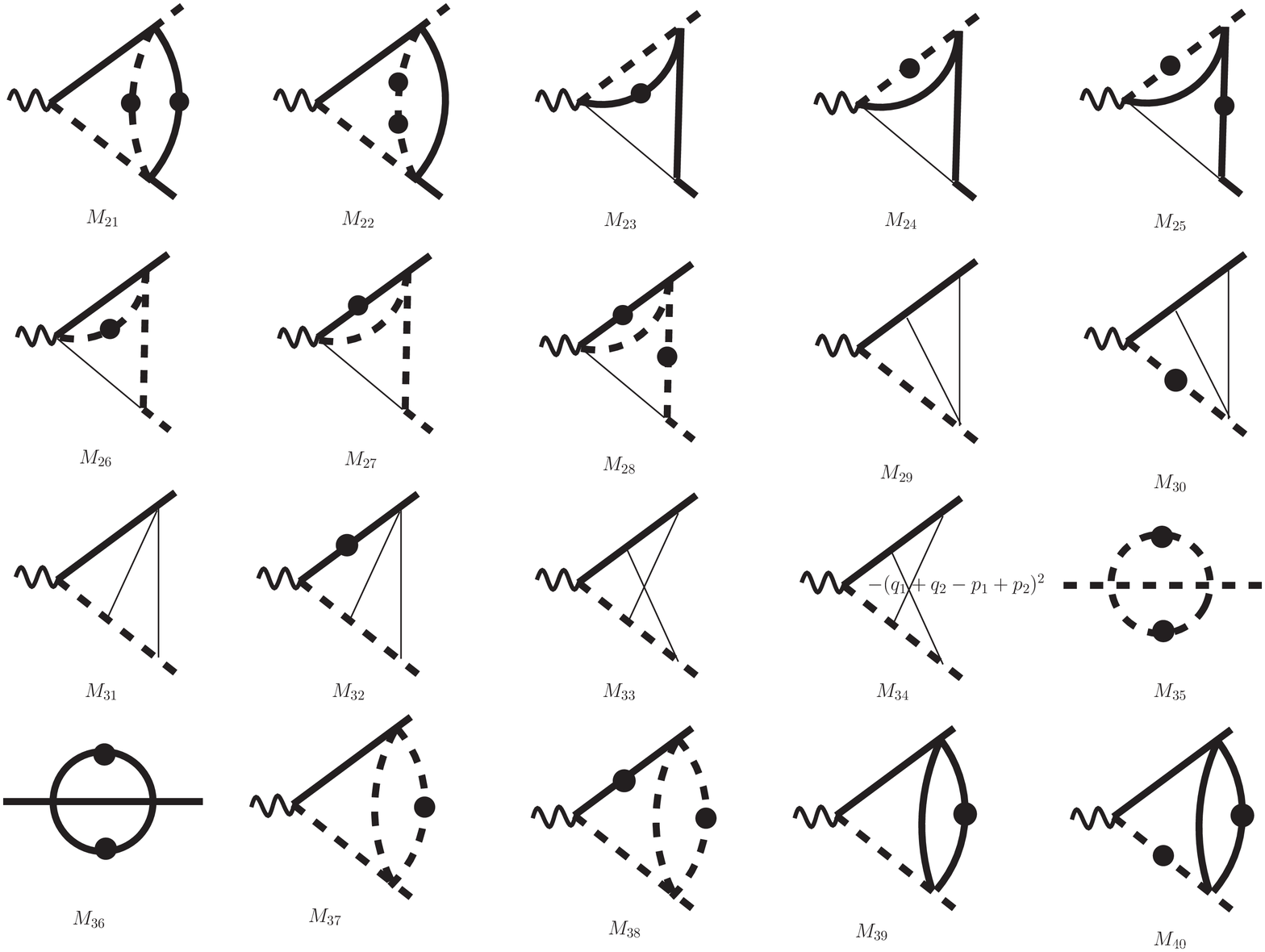}
\caption{All the NNLO master integrals for heavy-to-light form factors with different type of fermion masses. The thin lines denote massless propagators; the thick dashed lines denote massive fermion with mass $m_1$; and the thick solid lines denote massive fermion with mass $m_2$.  A dot on a propagator indicates that the power of the propagator is raised to 2. Two dots represents that the propagator is raised to power 3.}
\label{midiag}
\end{center}
\end{figure}

All the amplitudes of two-loop QCD or QED corrections for the processes we concern can be reduced to a set of 40 master integrals, including two non-planar integrals.  The master integrals $M_i (i=1\ldots 40)$ are shown in Fig. \ref{midiag}. The vector of canonical basis ${\bf F}$ is built up with 40 functions $F_i(s,m_1,m_2,\epsilon) (i=1\ldots 40)$, defined in terms of the linear combinations of master integrals $M_i$.

\bea
F_1 & = & \, \epsilon^2 \, M_1 \, , \label{f1}\\
F_2 & = & \, \epsilon^2 \, M_2 \, , \label{f2}\\
F_3 & = & \, \epsilon^2 \, M_3 \, , \label{f2}\\
F_4 & = & \, \epsilon^2 \, m_2^2\,   M_4\, ,\label{f4}\\
F_5 & = & \, \epsilon^2 \, m_1^2\,   M_5\, ,\label{f5}\\
F_6 & = & \, \epsilon^2 \, m_2^2\,   M_6\, ,\label{f6}\\
F_7 & = & \, \epsilon^2 \, m_1\, m_2\, (2 M_6 + M_7)\, ,\label{f7}\\
F_8 & = & \, \epsilon^2 \, m_1^2\,   M_8\, ,\label{f8}\\
F_9 & = & \, \epsilon^2 \, m_1\, m_2\, (2 M_8 + M_9)\, ,\label{f9}\\
F_{10} & = & \, \epsilon^2 \, s\,  M_{10} \, ,\label{f10}\\
F_{11} & = & \, \epsilon^2 \, \sqrt{s-(m_1+m_2)^2}\sqrt{s-(m_1-m_2)^2}\, (M_{10} + M_{11} + M_{12}) \, ,\label{f11}\\
F_{12} & = & \, \epsilon^2 \,\left( (m_1^2-m_2^2)\, (M_{10} + M_{11} + M_{12})\, + s\, (M_{11} - M_{12})\right)\, ,\label{f12}
\eea
\bea
F_{13} & = & \, \epsilon^2 \, \frac{\sqrt{s-(m_1+m_2)^2}\sqrt{s-(m_1-m_2)^2}}{2(s - m_1^2 + m_2^2)}\, (2 s\, M_{13} - M_{2} + M_{3}) \, ,\label{f13}\\
F_{14} & = & \, \epsilon^2 \, \frac{\sqrt{s-(m_1+m_2)^2}\sqrt{s-(m_1-m_2)^2}}{2(s + m_1^2 - m_2^2)}\, (2 s\, M_{14} - M_{1} + M_{3}) \, ,\label{f14}\\
F_{15} & = & \, \epsilon^2 \, \left( \frac{s^2(s-(m_1+m_2)^2)(s-(m_1-m_2)^2)}{(s+m_1^2-m_2^2)^2}M_{15} \right.\nonumber\\
& & \left. +\frac{\sqrt{s-(m_1+m_2)^2}\sqrt{s-(m_1-m_2)^2}}{s+m_1^2-m_2^2}\frac{(F_{13}-F_{14})}{\epsilon^2}\right.\nonumber\\
& & \left. + \frac{s\, m_1^2}{(s+m_1^2-m_2^2)^2}(M_1+M_2-2M_3) \right) \, ,\label{f15}\\
F_{16} & = & \, \epsilon^3 \, \sqrt{s-(m_1+m_2)^2}\sqrt{s-(m_1-m_2)^2}\, M_{16}  \, ,\label{f16}\\
F_{17} & = & \, \epsilon^3 \, \sqrt{s-(m_1+m_2)^2}\sqrt{s-(m_1-m_2)^2}\, M_{17}  \, ,\label{f17}\\
F_{18} & = & \, \epsilon^3 \, \sqrt{s-(m_1+m_2)^2}\sqrt{s-(m_1-m_2)^2}\, M_{18}  \, ,\label{f18}\\
F_{19} & = & \, \epsilon^3 \, \sqrt{s-(m_1+m_2)^2}\sqrt{s-(m_1-m_2)^2}\, M_{19}  \, ,\label{f19}\\
F_{20} & = & \, \epsilon^3 \, \sqrt{s-(m_1+m_2)^2}\sqrt{s-(m_1-m_2)^2}\, M_{20}  \, ,\label{f20}\\
F_{21} & = & \, \epsilon^2 \, \left(2s\,(\epsilon(M_{19}+2M_{20})-m_2^2 M_{21}-2m_1^2 M_{22})\right.\nonumber\\
& & \left.+2(m_2^2-m_1^2)(\epsilon(M_{19}+2M_{20})+m_2^2 M_{21}-2m_1^2 M_{22})\right)+2\frac{m_2}{m_1}F_9  \, ,\label{f21}\nonumber\\
F_{22} & = & \, \epsilon^2 \, \left( 2(m_1^2-m_2^2)(2\epsilon(M_{19}+2M_{20})+(m_1^2+m_2^2-s) M_{21}-4m_1^2 M_{22}) \right)\nonumber\\
& & +2\frac{m_1}{m_2}F_7-2\frac{m_2}{m_1}F_9\, ,\label{f22}\\
F_{23} & = & \, \epsilon^3 \, \sqrt{s-(m_1+m_2)^2}\sqrt{s-(m_1-m_2)^2}\, M_{23}  \, ,\label{f23}\\
F_{24} & = & \, \epsilon^3 \, \sqrt{s-(m_1+m_2)^2}\sqrt{s-(m_1-m_2)^2}\, M_{24}  \, ,\label{f24}\\
F_{25} & = & \, \epsilon^2 \, \frac{s\, m_2^2(s-(m_1+m_2)^2)(s-(m_1-m_2)^2)}{(s-m_1^2+m_2^2)^2}\, M_{25}\nonumber\\
& &\, + \epsilon^3 \, \frac{(s-(m_1+m_2)^2)(s-(m_1-m_2)^2)}{2(s-m_1^2+m_2^2)}\, (M_{23} - M_{24})\nonumber\\
& &\, -\frac{ m_2\, s\, (s-m_1^2-m_2^2)}{m_1(s-m_1^2+m_2^2)^2}F_9 -\frac{\sqrt{s-(m_1+m_2)^2}\sqrt{s-(m_1-m_2)^2}}{4(s-m_1^2+m_2^2)}F_{11}\nonumber\\
& &\, +\frac{s\, m_2^2}{(s-m_1^2+m_2^2)^2}(F_2-F_3-6F_8+2F_{12}) \, ,\label{f25}\\
F_{26} & = & \, \epsilon^3 \, \sqrt{s-(m_1+m_2)^2}\sqrt{s-(m_1-m_2)^2}\, M_{26}  \, ,\label{f26}\\
F_{27} & = & \, \epsilon^3 \, \sqrt{s-(m_1+m_2)^2}\sqrt{s-(m_1-m_2)^2}\, M_{27}  \, ,\label{f27}\\
F_{28} & = & \, \epsilon^2 \, \frac{s\, m_1^2(s-(m_1+m_2)^2)(s-(m_1-m_2)^2)}{(s+m_1^2-m_2^2)^2}\, M_{28}\nonumber\\
& &\, + \epsilon^3 \, \frac{(s-(m_1+m_2)^2)(s-(m_1-m_2)^2)}{2(s+m_1^2-m_2^2)}\, (M_{26} - M_{27})\nonumber\\
& &\, -\frac{ m_1\, s\, (s-m_1^2-m_2^2)}{m_2(s+m_1^2-m_2^2)^2}F_7 -\frac{\sqrt{s-(m_1+m_2)^2}\sqrt{s-(m_1-m_2)^2}}{4(s+m_1^2-m_2^2)}F_{11}\nonumber\\
& &\, +\frac{s\, m_1^2}{(s+m_1^2-m_2^2)^2}(F_1-F_3-6F_6-2F_{12}) \, ,\label{f28}
\eea

\bea
F_{29} & = & \, \epsilon^4 \, \sqrt{s-(m_1+m_2)^2}\sqrt{s-(m_1-m_2)^2}\, M_{29}  \, ,\label{f29}\\
F_{30} & = & \, \epsilon^3 \, (s-(m_1+m_2)^2)(s-(m_1-m_2)^2)\, M_{30}  \, ,\label{f30}\\
F_{31} & = & \, \epsilon^4 \, \sqrt{s-(m_1+m_2)^2}\sqrt{s-(m_1-m_2)^2}\, M_{31}  \, ,\label{f31}\\
F_{32} & = & \, \epsilon^3 \, (s-(m_1+m_2)^2)(s-(m_1-m_2)^2)\, M_{32}  \, ,\label{f32}\\
F_{33} & = & \, \epsilon^4 \, (s-(m_1+m_2)^2)(s-(m_1-m_2)^2)\, M_{33}  \, ,\label{f33}\\
F_{34} & = & \, \epsilon^4 \, \sqrt{s-(m_1+m_2)^2}\sqrt{s-(m_1-m_2)^2}\, M_{34}  \, ,\label{f34}\\
F_{35} & = & \, \epsilon^2 \, m_1^2\, M_{35}  \, ,\label{f35}\\
F_{36} & = & \, \epsilon^2 \, m_2^2\, M_{36}  \, ,\label{f36}\\
F_{37} & = & \, \epsilon^3 \, \sqrt{s-(m_1+m_2)^2}\sqrt{s-(m_1-m_2)^2}\, M_{37}  \, ,\label{f37}\\
F_{38} & = & \, \epsilon^2 \, (s-(m_1+m_2)^2)(s-(m_1-m_2)^2)\, M_{38} + \frac{s-m_1^2-m_2^2}{2 m_1\, m_2}F_7   \, ,\label{f38}\\
F_{39} & = & \, \epsilon^3 \, \sqrt{s-(m_1+m_2)^2}\sqrt{s-(m_1-m_2)^2}\, M_{39}  \, ,\label{f39}\\
F_{40} & = & \, \epsilon^2 \, (s-(m_1+m_2)^2)(s-(m_1-m_2)^2)\, M_{40} + \frac{s-m_1^2-m_2^2}{2 m_1\, m_2}F_9   \, .\label{f40}\\
\eea

For our choice of master integrals above, the differential equations for $\text{{\bf F}}=(F_1\ldots F_{34})$  have the following canonical form
\bea
\text{d} \text{{\bf F}}(x,y;\epsilon)=\epsilon\, \text{d}\, \tilde{A}(x,y)\,  \text{{\bf F}}(x,y;\epsilon),
\eea
with
\bea
\tilde{A}(x,y) &=& A_1\, \ln(x) + A_2\, \ln(x+1) + A_3\, \ln(x-1)+ A_4\, \ln(x+y)+ A_5\, \ln(x-y)\nonumber\\
& & + A_6\, \ln(x\, y+1)+ A_7\, \ln(x\, y-1)+A_8\, \ln(y) + A_9\, \ln(y+1) + A_{10}\, \ln(y-1)\nonumber\\
& & + A_{11}\, \ln(x^2\, y-2 x+y) +A_{12}\, \ln(x^2-2 y\, x+1) .
\label{dlog}
\eea
The notations $A_i (i=1\cdots 12)$ are $40\times 40$ matrices with rational numbers, they are presented in appendix. A. For reader's convenience, the rational matrices in electronic form are also provided in ancillary file accompanying the \textbf{arXiv} version of this paper. We can see in equation (\ref{dlog}) that in equal masses
case with $y=1$, the alphabet turns into $\{x,x+1,x-1\}$, and the results of all the canonical basis can simply be reexpressed in terms of Harmonic polylogarithms.

The integral $M_1$ is defined as follow
\be
M_1 = \int {\mathcal D}^Dq_1 \, {\mathcal D}^Dq_2 \, \frac{1}{(-q_1^2+m_1^2)^2} \, \frac{1}{(-q_2^2+m_1^2)^2} = \frac{1}{\epsilon^2} \, ,
\ee
with the measure of the integration is defined as
\be
{\mathcal D}^Dq_i = \frac{1}{\pi^{D/2}\Gamma(1+\epsilon)}\left(\frac{m_1^2}{\mu^2}\right)^\epsilon  d^Dq_i \, .
\ee
For master integrals without numerators, their definition can readily be read off from Fig. \ref{midiag}, with the normalization defined above. For master integrals with numerator, we first define a series of propagators
\begin{align}
P_{1} & =m_{2}^{2}-q_{1}^{2},\hspace{3.1cm}P_{2}=m_{1}^{2}-q_{2}^{2},\nonumber \\
P_{3} & =-(q_{1}-p_{1})^{2},\hspace{2.52cm}P_{4}=-(q_{2}+p_{2})^{2},\nonumber \\
P_{5} & =m_{2}^{2}-(q_{1}+q_{2}+p_{2})^{2},\hspace{1.03cm}P_{6}=m_{1}^{2}-(q_{1}+q_{2}-p_{1})^{2},\nonumber \\
P_{7} & =-(q_{1}+q_{2}-p_{1}+p_{2})^{2}.
\end{align}
Then, the master integral with numerator $(M_{34})$ can be expressed as
\bea
M_{34} & = & \int {\mathcal D}^Dq_1 \, {\mathcal D}^Dq_2 \frac{P_7}{P_1 P_2 P_3 P_4 P_5 P_6}.
\eea

\section{Boundary conditions and analytic continuation}

Now, we are ready to perform the calculations of the differential equations. The first step is to specify all the boundary conditions that will completely fix the solutions of the differential equations. The results of basis $(F_1\ldots F_5)$ and $(F_{35},F_{36})$ are already known in the literature \cite{Henn:2013woa, Chen:2015csa,remid, Chen:2017xqd}. They can be recalculated with the assistance of Mathematica packages {\bf MB} \cite{Czakon:2005rk} and {\bf AMBRE} \cite{Gluza:2007rt,Gluza:2010rn,Blumlein:2014maa}. The integrals $(F_6 \ldots F_9)$ have been calculated in \cite{Chen:2015csa, remid}, their boundary conditions could be determined by setting $y=\frac{m_2}{m_1}=1$, with the masses of two fermions equaling to each other, and their boundaries are proportional to the results of $F_{35}$.

The integral $M_{10}$ does not have singularity at $s=0$. Thanks to the normalization factor $s$ that multiplying with $M_{10}$ in $F_{10}$, we can readily know that $F_{10}=0$ at $s=0$.
Considering the fact that $M_{(11,13,14,16\ldots 20,23,24,26,27,29\ldots 34,37,39)}$ are regular at $s=(m_1-m_2)^2$ as well as $s=(m_1+m_2)^2$, and their normalization factor to be $\sqrt{s-(m_1+m_2)^2}\sqrt{s-(m_1-m_2)^2}$ or $(s-(m_1+m_2)^2)(s-(m_1-m_2)^2)$, the boundaries of basis $F_{(11,13,14,16\ldots 20,23,24,26,27,29\ldots 34,37,39)}$ are  0 at $s=(m_1-m_2)^2$ and $s=(m_1+m_2)^2$. The boundary of $F_{15}$ at $s=(m_1-m_2)^2$ and $s=(m_1+m_2)^2$ can also be determined similarly and then expressed as
\bea
F_{15}|_{s=\{(m_1-m_2)^2,(m_1+m_2)^2\}}&=&\frac{1}{4}(F_1+F_2-2F_3)|_{s=\{(m_1-m_2)^2,(m_1+m_2)^2\}}.
\eea

Considering the fact that $M_{12}$ does not have singularity at $s = 0$, \ie $(x=y)$ or $(x=\frac{1}{y})$ , the boundary condition of $F_{12}$ can be determined from the
differential equation of $F_{12}$. To further illustrate it, we consider the differential of $F_{12}$ with respect to variable $x$ and find that
\be
\frac{\partial F_{12}}{\partial x}=\epsilon\, \left(-\frac{F_{11}+F_{12}}{x-y} + y\, \frac{F_{11}-F_{12}}{x\, y-1}+\frac{F_{12}}{x}\right).
\ee
Both $F_{11}$ and $F_{12}$ have finite limit at $x=y$ and $x=\frac{1}{y}$, this consistency leads to two relations between $F_{11}$ and $F_{12}$
\bea
F_{11}|_{x=y} &=& -F_{12}|_{x=y}.\nonumber\\
F_{11}|_{x=\frac{1}{y}} &=& F_{12}|_{x=\frac{1}{y}}.
\eea
The boundary of $(F_{21},F_{22},F_{25},F_{28},F_{38},F_{40})$ can also be determined with the discussion above. By now, all the boundary conditions are fixed.

We then proceed to determinate the analytic continuation of all the master integrals. When considering the processes (\ref{pro1},\ref{pro2}), the variables of the master integrals lie either in Euclidean region or in Minkowski region. In order to obtain the correct numerical results, their analytic continuation should be considered carefully. The proper analytic continuation can be achieved by the replacement $s \rightarrow s + i 0$ at fixed $m_1^2$ and $m_2^2$. This transfer corresponds to $x\rightarrow x + i 0$.

After the determination of the boundary conditions and analytical continuation, we can readily obtain the analytical results of all the master integrals using the differential equations we obtain. The results of the master integrals can be written in terms of Goncharov polylogarithms introduced in section 3. The results for $F_i(i=1\ldots40)$ are calculated up to weight four in this work. All the analytic results up to weight four are collected in the ancillary file "{\bf results.m}" which is supplied with the paper. Here, for illustration, we present the results for integrals $(F_6,F_{10},F_{33})$  up to weight three.
\bea
F_6 &=& \epsilon^2\, (G_{-1,0}(y)+G_{1,0}(y))-2\epsilon^3\, (G_{1,1,0}(y)+G_{-1,-1,0}(y)+G_{1,0,0}(y)+G_{-1,0,0}(y)\nonumber\\
& & -G_{0,1,0}(y)-G_{0,-1,0}(y)+2G_{1,-1,0}(y)+2G_{-1,1,0}(y))+{\cal O}(\epsilon^{4}),\nonumber\\
F_{10} &=& 2\epsilon^2\, \big[G_{0,0}(y)-G_{0,0}(x)\big]+\epsilon^3\, \big[\frac{G_{0,0}(y)}{2}(G_{\frac{1}{y}}(x)+G_{y}(x)+4G_{y}(1)-4G_{\frac{1}{y}}(1)-G_{\frac{1}{y}}(y))\nonumber\\
&+& 4G_0(y)(G_{0,y}(x)-G_{0,\frac{1}{y}}(x)+G_{0,0}(x)+G_{0,\frac{1}{y}}(y)-G_{\frac{1}{y},0}(1)-G_{y,0}(1)+\frac{\pi^2}{3})\nonumber\\
&+& G_0(x)(\frac{G_{0,0}(y)}{2}-4(G_{\frac{1}{y}}(1)-G_{y}(1))G_0(y)-4G_{\frac{1}{y},0}(1)-4G_{y,0}(1)+\pi^2)-6G_{0,0,0}(x) \nonumber\\
&+& 12(G_{0,-1,0}(x)+G_{0,1,0}(x)-G_{0,-1,0}(y)-G_{0,1,0}(y))-2(2G_{0,\frac{1}{y},0}(x)+2G_{0,y,0}(x)\nonumber\\
&+& G_{\frac{1}{y},0,0}(x)+G_{y,0,0}(x))+2G_{\frac{1}{y},0,0}(y)+4G_{0,\frac{1}{y},0}(y)+6\zeta(3)\big]+{\cal O}(\epsilon^{4}),\nonumber\\
F_{33} &=& \epsilon^3\, \big[2G_{0,0,0}(x)-2G_{0,1,0}(x)-2G_{0,-1,0}(x)+\frac{1}{6}\pi^2 G_0(x)+\zeta(3)\big]+{\cal O}(\epsilon^{4}).
\eea
Note that the weight three results of $F_{33}$ depend only on variable $x$, the  dependence of $F_{33}$ on variable $y$ will start at ${\cal O}(\epsilon^{4})$.

The calculations are performed with our in house Mathematica code. It is desirable to perform an independent check on the analytical expressions of the master integrals. We check all the analytical results against the numerical results obtained from packages {\bf Fiesta} \cite{Smirnov:2013eza,Smirnov:2015mct} and {\bf SecDec} \cite{Borowka:2012yc,Borowka:2015mxa}. Perfect agreement has been achieved between the analytical and numerical approaches with kinematics in both Euclidean region and Minkowski region. To be more specific, we show the numerical results of the master integrals $M_{33}(s,m_1,m_2)$ obtained from packages {\bf Fiesta} and {\bf SecDec}  and our analytical expressions in two different kinematics,
\bea
M_{33}^{\text{SecDec}}(5.4,1.0,0.2)&=& \frac{0.890536+i1.307051 \pm (0.000010+i0.000012)}{\epsilon} \nonumber\\
& & -5.989791+i7.770764 \pm(0.000306+ i0.000385), \nonumber\\
M_{33}^{\text{FIESTA}}(5.4,1.0,0.2)&=& \frac{0.890545 + i1.307058 \pm (0.000036  + i0.000039)}{\epsilon}\nonumber\\
& & -5.989754 + i7.772445 \pm(0.00064 + i0.000639),\nonumber\\
M_{33}^{\text{Ours}}(5.4,1.0,0.2)&=& \frac{0.8905387485\ldots + i1.307054048\ldots}{\epsilon}\nonumber\\
& &-5.989494850\ldots + i7.772448208\ldots,\nonumber\\ \nonumber\\
M_{33}^{\text{SecDec}}(-5.4,1.0,0.2)&=& \frac{-0.4466129 \pm 0.0000004 }{\epsilon} -0.507366 \pm 0.000006  , \nonumber\\
M_{33}^{\text{FIESTA}}(-5.4,1.0,0.2)&=& \frac{-0.446613 \pm 0.000005 }{\epsilon} -0.507387 \pm 0.000049  ,\nonumber\\
M_{33}^{\text{Ours}}(-5.4,1.0,0.2)&=& \frac{-0.4466129967\ldots }{\epsilon}- 0.5073683817\ldots .
\eea
We can see from the equations above that the numerical results obtained from packages {\bf Fiesta} and {\bf SecDec} with error estimates are perfectly agree to that precision with the numerical results obtained from our analytical answer. Note that it took {\bf Fiesta} and {\bf SecDec} packages each more than ten hours on a workstation with 16 cores (2.3 GHz) processor to reach the above precision for kinematic $(s=5.4,m_1=1.0,m_2=0.2)$ in Minkowski region.

\section{Discussions and Conclusions}

In summary, applying the method of differential equations, we calculate the full set of the two-loop master integrals for heavy-to-light form factors of two different types of massive fermions and arbitrary momentum transfer in NNLO QCD or QED corrections. With a proper choice of master integrals, it turns out that we can cast all the differential equations into the canonical form, which can straightforwardly be integrated order by order in dimensional regularization parameter $\epsilon$. Under the determination of all boundary conditions, we express the results of all the basis in terms of Goncharov polylogarithms. The integrals allow to determine the two-loop amplitudes for heavy-to-light form factors of two massive fermions in a full analytical way, and thus to compute the NNLO corrections to the decay of heavy massive fermion into light massive fermion, or the decay of $W$ boson into two massive quarks without any approximations. The integrals are also applicable to the calculations of NNLO QCD corrections to the inclusive decay of B meson into charmonium, and two-loop form factors of $H^+t\bar{b}$ coupling vertex.

\acknowledgments

This work was supported by the National Natural Science
Foundation of China (NSFC) under the grants 11747051.

\appendix


\section{The Matrices R}
The rational matrices $A_i(i=1\ldots 10)$ are expressed as
\begin{align}
 A_{1} &=\scalemath{0.47}{
  \left(
\begin{array}{cccccccccccccccccccccccccccccccccccccccc}
\pminus  0 & \pminus  0 & \pminus  0 & \pminus  0 & \pminus  0 & \pminus  0 & \pminus  0 & \pminus  0 & \pminus  0 & \pminus  0 & \pminus  0 & \pminus  0 & \pminus  0 & \pminus  0 & \pminus  0 & \pminus  0 & \pminus  0 & \pminus  0 & \pminus  0 & \pminus  0 & \pminus  0 & \pminus  0 & \pminus  0 & \pminus  0 & \pminus  0 & \pminus  0 & \pminus  0 & \pminus  0 & \pminus  0 & \pminus  0 & \pminus  0 & \pminus  0 & \pminus  0 & \pminus  0 & \pminus  0 & \pminus  0 & \pminus  0 & \pminus  0 & \pminus  0 & \pminus  0 \\ \pminus
 0 & \pminus  0 & \pminus  0 & \pminus  0 & \pminus  0 & \pminus  0 & \pminus  0 & \pminus  0 & \pminus  0 & \pminus  0 & \pminus  0 & \pminus  0 & \pminus  0 & \pminus  0 & \pminus  0 & \pminus  0 & \pminus  0 & \pminus  0 & \pminus  0 & \pminus  0 & \pminus  0 & \pminus  0 & \pminus  0 & \pminus  0 & \pminus  0 & \pminus  0 & \pminus  0 & \pminus  0 & \pminus  0 & \pminus  0 & \pminus  0 & \pminus  0 & \pminus  0 & \pminus  0 & \pminus  0 & \pminus  0 & \pminus  0 & \pminus  0 & \pminus  0 & \pminus  0 \\ \pminus
 0 & \pminus  0 & \pminus  0 & \pminus  0 & \pminus  0 & \pminus  0 & \pminus  0 & \pminus  0 & \pminus  0 & \pminus  0 & \pminus  0 & \pminus  0 & \pminus  0 & \pminus  0 & \pminus  0 & \pminus  0 & \pminus  0 & \pminus  0 & \pminus  0 & \pminus  0 & \pminus  0 & \pminus  0 & \pminus  0 & \pminus  0 & \pminus  0 & \pminus  0 & \pminus  0 & \pminus  0 & \pminus  0 & \pminus  0 & \pminus  0 & \pminus  0 & \pminus  0 & \pminus  0 & \pminus  0 & \pminus  0 & \pminus  0 & \pminus  0 & \pminus  0 & \pminus  0 \\
\end{array}
\right)
  },
\end{align}
\begin{align}
 A_{2} &=\scalemath{0.48}{
  \left(
\begin{array}{cccccccccccccccccccccccccccccccccccccccc}
\pminus 0 & \pminus 0 & \pminus 0 & \pminus 0 & \pminus 0 & \pminus 0 & \pminus 0 & \pminus 0 & \pminus 0 & \pminus 0 & \pminus 0 & \pminus 0 & \pminus 0 & \pminus 0 & \pminus 0 & \pminus 0 & \pminus 0 & \pminus 0 & \pminus 0 & \pminus 0 & \pminus 0 & \pminus 0 & \pminus 0 & \pminus 0 & \pminus 0 & \pminus 0 & \pminus 0 & \pminus 0 & \pminus 0 & \pminus 0 & \pminus 0 & \pminus 0 & \pminus 0 & \pminus 0 & \pminus 0 & \pminus 0 & \pminus 0 & \pminus 0 & \pminus 0 & \pminus 0 \\ \pminus
 0 & \pminus 0 & \pminus 0 & \pminus 0 & \pminus 0 & \pminus 0 & \pminus 0 & \pminus 0 & \pminus 0 & \pminus 0 & \pminus 0 & \pminus 0 & \pminus 0 & \pminus 0 & \pminus 0 & \pminus 0 & \pminus 0 & \pminus 0 & \pminus 0 & \pminus 0 & \pminus 0 & \pminus 0 & \pminus 0 & \pminus 0 & \pminus 0 & \pminus 0 & \pminus 0 & \pminus 0 & \pminus 0 & \pminus 0 & \pminus 0 & \pminus 0 & \pminus 0 & \pminus 0 & \pminus 0 & \pminus 0 & \pminus 0 & \pminus 0 & \pminus 0 & \pminus 0 \\ \pminus
 0 & \pminus 0 & \pminus 0 & \pminus 0 & \pminus 0 & \pminus 0 & \pminus 0 & \pminus 0 & \pminus 0 & \pminus 0 & \pminus 0 & \pminus 0 & \pminus 0 & \pminus 0 & \pminus 0 & \pminus 0 & \pminus 0 & \pminus 0 & \pminus 0 & \pminus 0 & \pminus 0 & \pminus 0 & \pminus 0 & \pminus 0 & \pminus 0 & \pminus 0 & \pminus 0 & \pminus 0 & \pminus 0 & \pminus 0 & \pminus 0 & \pminus 0 & \pminus 0 & \pminus 0 & \pminus 0 & \pminus 0 & \pminus 0 & \pminus 0 & \pminus 0 & \pminus 0 \\ \pminus
 0 & \pminus 0 & \pminus 0 & \pminus 0 & \pminus 0 & \pminus 0 & \pminus 0 & \pminus 0 & \pminus 0 & \pminus 0 & \pminus 0 & \pminus 0 & \pminus 0 & \pminus 0 & \pminus 0 & \pminus 0 & \pminus 0 & \pminus 0 & \pminus 0 & \pminus 0 & \pminus 0 & \pminus 0 & \pminus 0 & \pminus 0 & \pminus 0 & \pminus 0 & \pminus 0 & \pminus 0 & \pminus 0 & \pminus 0 & \pminus 0 & \pminus 0 & \pminus 0 & \pminus 0 & \pminus 0 & \pminus 0 & \pminus 0 & \pminus 0 & \pminus 0 & \pminus 0 \\
\end{array}
\right)
  },
\end{align}
\begin{align}
 A_{3} &=\scalemath{0.5}{
  \left(
\begin{array}{cccccccccccccccccccccccccccccccccccccccc}
\pminus 0 & \pminus 0 & \pminus 0 & \pminus 0 & \pminus 0 & \pminus 0 & \pminus 0 & \pminus 0 & \pminus 0 & \pminus 0 & \pminus 0 & \pminus 0 & \pminus 0 & \pminus 0 & \pminus 0 & \pminus 0 & \pminus 0 & \pminus 0 & \pminus 0 & \pminus 0 & \pminus 0 & \pminus 0 & \pminus 0 & \pminus 0 & \pminus 0 & \pminus 0 & \pminus 0 & \pminus 0 & \pminus 0 & \pminus 0 & \pminus 0 & \pminus 0 & \pminus 0 & \pminus 0 & \pminus 0 & \pminus 0 & \pminus 0 & \pminus 0 & \pminus 0 & \pminus 0 \\ \pminus
 0 & \pminus 0 & \pminus 0 & \pminus 0 & \pminus 0 & \pminus 0 & \pminus 0 & \pminus 0 & \pminus 0 & \pminus 0 & \pminus 0 & \pminus 0 & \pminus 0 & \pminus 0 & \pminus 0 & \pminus 0 & \pminus 0 & \pminus 0 & \pminus 0 & \pminus 0 & \pminus 0 & \pminus 0 & \pminus 0 & \pminus 0 & \pminus 0 & \pminus 0 & \pminus 0 & \pminus 0 & \pminus 0 & \pminus 0 & \pminus 0 & \pminus 0 & \pminus 0 & \pminus 0 & \pminus 0 & \pminus 0 & \pminus 0 & \pminus 0 & \pminus 0 & \pminus 0 \\ \pminus
 0 & \pminus 0 & \pminus 0 & \pminus 0 & \pminus 0 & \pminus 0 & \pminus 0 & \pminus 0 & \pminus 0 & \pminus 0 & \pminus 0 & \pminus 0 & \pminus 0 & \pminus 0 & \pminus 0 & \pminus 0 & \pminus 0 & \pminus 0 & \pminus 0 & \pminus 0 & \pminus 0 & \pminus 0 & \pminus 0 & \pminus 0 & \pminus 0 & \pminus 0 & \pminus 0 & \pminus 0 & \pminus 0 & \pminus 0 & \pminus 0 & \pminus 0 & \pminus 0 & \pminus 0 & \pminus 0 & \pminus 0 & \pminus 0 & \pminus 0 & \pminus 0 & \pminus 0 \\ \pminus
 0 & \pminus 0 & \pminus 0 & \pminus 0 & \pminus 0 & \pminus 0 & \pminus 0 & \pminus 0 & \pminus 0 & \pminus 0 & \pminus 0 & \pminus 0 & \pminus 0 & \pminus 0 & \pminus 0 & \pminus 0 & \pminus 0 & \pminus 0 & \pminus 0 & \pminus 0 & \pminus 0 & \pminus 0 & \pminus 0 & \pminus 0 & \pminus 0 & \pminus 0 & \pminus 0 & \pminus 0 & \pminus 0 & \pminus 0 & \pminus 0 & \pminus 0 & \pminus 0 & \pminus 0 & \pminus 0 & \pminus 0 & \pminus 0 & \pminus 0 & \pminus 0 & \pminus 0 \\
\end{array}
\right)
  },
\end{align}
\begin{align}
 A_{4} &=\scalemath{0.5}{
  \left(
\begin{array}{cccccccccccccccccccccccccccccccccccccccc}
\pminus 0 & \pminus 0 & \pminus 0 & \pminus 0 & \pminus 0 & \pminus 0 & \pminus 0 & \pminus 0 & \pminus 0 & \pminus 0 & \pminus 0 & \pminus 0 & \pminus 0 & \pminus 0 & \pminus 0 & \pminus 0 & \pminus 0 & \pminus 0 & \pminus 0 & \pminus 0 & \pminus 0 & \pminus 0 & \pminus 0 & \pminus 0 & \pminus 0 & \pminus 0 & \pminus 0 & \pminus 0 & \pminus 0 & \pminus 0 & \pminus 0 & \pminus 0 & \pminus 0 & \pminus 0 & \pminus 0 & \pminus 0 & \pminus 0 & \pminus 0 & \pminus 0 & \pminus 0 \\ \pminus
 0 & \pminus 0 & \pminus 0 & \pminus 0 & \pminus 0 & \pminus 0 & \pminus 0 & \pminus 0 & \pminus 0 & \pminus 0 & \pminus 0 & \pminus 0 & \pminus 0 & \pminus 0 & \pminus 0 & \pminus 0 & \pminus 0 & \pminus 0 & \pminus 0 & \pminus 0 & \pminus 0 & \pminus 0 & \pminus 0 & \pminus 0 & \pminus 0 & \pminus 0 & \pminus 0 & \pminus 0 & \pminus 0 & \pminus 0 & \pminus 0 & \pminus 0 & \pminus 0 & \pminus 0 & \pminus 0 & \pminus 0 & \pminus 0 & \pminus 0 & \pminus 0 & \pminus 0 \\ \pminus
 0 & \pminus 0 & \pminus 0 & \pminus 0 & \pminus 0 & \pminus 0 & \pminus 0 & \pminus 0 & \pminus 0 & \pminus 0 & \pminus 0 & \pminus 0 & \pminus 0 & \pminus 0 & \pminus 0 & \pminus 0 & \pminus 0 & \pminus 0 & \pminus 0 & \pminus 0 & \pminus 0 & \pminus 0 & \pminus 0 & \pminus 0 & \pminus 0 & \pminus 0 & \pminus 0 & \pminus 0 & \pminus 0 & \pminus 0 & \pminus 0 & \pminus 0 & \pminus 0 & \pminus 0 & \pminus 0 & \pminus 0 & \pminus 0 & \pminus 0 & \pminus 0 & \pminus 0 \\ \pminus
 0 & \pminus 0 & \pminus 0 & \pminus 0 & \pminus 0 & \pminus 0 & \pminus 0 & \pminus 0 & \pminus 0 & \pminus 0 & \pminus 0 & \pminus 0 & \pminus 0 & \pminus 0 & \pminus 0 & \pminus 0 & \pminus 0 & \pminus 0 & \pminus 0 & \pminus 0 & \pminus 0 & \pminus 0 & \pminus 0 & \pminus 0 & \pminus 0 & \pminus 0 & \pminus 0 & \pminus 0 & \pminus 0 & \pminus 0 & \pminus 0 & \pminus 0 & \pminus 0 & \pminus 0 & \pminus 0 & \pminus 0 & \pminus 0 & \pminus 0 & \pminus 0 & \pminus 0 \\
\end{array}
\right)
  },
\end{align}
\begin{align}
 A_{5} &=\scalemath{0.5}{
  \left(
\begin{array}{cccccccccccccccccccccccccccccccccccccccc}
\pminus 0 & \pminus  0 & \pminus  0 & \pminus  0 & \pminus  0 & \pminus  0 & \pminus  0 & \pminus  0 & \pminus  0 & \pminus  0 & \pminus  0 & \pminus  0 & \pminus  0 & \pminus  0 & \pminus  0 & \pminus  0 & \pminus  0 & \pminus  0 & \pminus  0 & \pminus  0 & \pminus  0 & \pminus  0 & \pminus  0 & \pminus  0 & \pminus  0 & \pminus  0 & \pminus  0 & \pminus  0 & \pminus  0 & \pminus  0 & \pminus  0 & \pminus  0 & \pminus  0 & \pminus  0 & \pminus  0 & \pminus  0 & \pminus  0 & \pminus  0 & \pminus  0 & \pminus  0 \\ \pminus
 0 & \pminus  0 & \pminus  0 & \pminus  0 & \pminus  0 & \pminus  0 & \pminus  0 & \pminus  0 & \pminus  0 & \pminus  0 & \pminus  0 & \pminus  0 & \pminus  0 & \pminus  0 & \pminus  0 & \pminus  0 & \pminus  0 & \pminus  0 & \pminus  0 & \pminus  0 & \pminus  0 & \pminus  0 & \pminus  0 & \pminus  0 & \pminus  0 & \pminus  0 & \pminus  0 & \pminus  0 & \pminus  0 & \pminus  0 & \pminus  0 & \pminus  0 & \pminus  0 & \pminus  0 & \pminus  0 & \pminus  0 & \pminus  0 & \pminus  0 & \pminus  0 & \pminus  0 \\ \pminus
 0 & \pminus  0 & \pminus  0 & \pminus  0 & \pminus  0 & \pminus  0 & \pminus  0 & \pminus  0 & \pminus  0 & \pminus  0 & \pminus  0 & \pminus  0 & \pminus  0 & \pminus  0 & \pminus  0 & \pminus  0 & \pminus  0 & \pminus  0 & \pminus  0 & \pminus  0 & \pminus  0 & \pminus  0 & \pminus  0 & \pminus  0 & \pminus  0 & \pminus  0 & \pminus  0 & \pminus  0 & \pminus  0 & \pminus  0 & \pminus  0 & \pminus  0 & \pminus  0 & \pminus  0 & \pminus  0 & \pminus  0 & \pminus  0 & \pminus  0 & \pminus  0 & \pminus  0 \\
\end{array}
\right)
  },
\end{align}
\begin{align}
 A_{6} &=\scalemath{0.5}{
  \left(
\begin{array}{cccccccccccccccccccccccccccccccccccccccc}
\pminus 0 & \pminus  0 & \pminus  0 & \pminus  0 & \pminus  0 & \pminus  0 & \pminus  0 & \pminus  0 & \pminus  0 & \pminus  0 & \pminus  0 & \pminus  0 & \pminus  0 & \pminus  0 & \pminus  0 & \pminus  0 & \pminus  0 & \pminus  0 & \pminus  0 & \pminus  0 & \pminus  0 & \pminus  0 & \pminus  0 & \pminus  0 & \pminus  0 & \pminus  0 & \pminus  0 & \pminus  0 & \pminus  0 & \pminus  0 & \pminus  0 & \pminus  0 & \pminus  0 & \pminus  0 & \pminus  0 & \pminus  0 & \pminus  0 & \pminus  0 & \pminus  0 & \pminus  0 \\ \pminus
 0 & \pminus  0 & \pminus  0 & \pminus  0 & \pminus  0 & \pminus  0 & \pminus  0 & \pminus  0 & \pminus  0 & \pminus  0 & \pminus  0 & \pminus  0 & \pminus  0 & \pminus  0 & \pminus  0 & \pminus  0 & \pminus  0 & \pminus  0 & \pminus  0 & \pminus  0 & \pminus  0 & \pminus  0 & \pminus  0 & \pminus  0 & \pminus  0 & \pminus  0 & \pminus  0 & \pminus  0 & \pminus  0 & \pminus  0 & \pminus  0 & \pminus  0 & \pminus  0 & \pminus  0 & \pminus  0 & \pminus  0 & \pminus  0 & \pminus  0 & \pminus  0 & \pminus  0 \\ \pminus
 0 & \pminus  0 & \pminus  0 & \pminus  0 & \pminus  0 & \pminus  0 & \pminus  0 & \pminus  0 & \pminus  0 & \pminus  0 & \pminus  0 & \pminus  0 & \pminus  0 & \pminus  0 & \pminus  0 & \pminus  0 & \pminus  0 & \pminus  0 & \pminus  0 & \pminus  0 & \pminus  0 & \pminus  0 & \pminus  0 & \pminus  0 & \pminus  0 & \pminus  0 & \pminus  0 & \pminus  0 & \pminus  0 & \pminus  0 & \pminus  0 & \pminus  0 & \pminus  0 & \pminus  0 & \pminus  0 & \pminus  0 & \pminus  0 & \pminus  0 & \pminus  0 & \pminus  0 \\
\end{array}
\right)
  },
\end{align}
\begin{align}
 A_{7} &=\scalemath{0.5}{
  \left(
\begin{array}{cccccccccccccccccccccccccccccccccccccccc}
\pminus  0 & \pminus  0 & \pminus  0 & \pminus  0 & \pminus  0 & \pminus  0 & \pminus  0 & \pminus  0 & \pminus  0 & \pminus  0 & \pminus  0 & \pminus  0 & \pminus  0 & \pminus  0 & \pminus  0 & \pminus  0 & \pminus  0 & \pminus  0 & \pminus  0 & \pminus  0 & \pminus  0 & \pminus  0 & \pminus  0 & \pminus  0 & \pminus  0 & \pminus  0 & \pminus  0 & \pminus  0 & \pminus  0 & \pminus  0 & \pminus  0 & \pminus  0 & \pminus  0 & \pminus  0 & \pminus  0 & \pminus  0 & \pminus  0 & \pminus  0 & \pminus  0 & \pminus  0 \\ \pminus
 0 & \pminus  0 & \pminus  0 & \pminus  0 & \pminus  0 & \pminus  0 & \pminus  0 & \pminus  0 & \pminus  0 & \pminus  0 & \pminus  0 & \pminus  0 & \pminus  0 & \pminus  0 & \pminus  0 & \pminus  0 & \pminus  0 & \pminus  0 & \pminus  0 & \pminus  0 & \pminus  0 & \pminus  0 & \pminus  0 & \pminus  0 & \pminus  0 & \pminus  0 & \pminus  0 & \pminus  0 & \pminus  0 & \pminus  0 & \pminus  0 & \pminus  0 & \pminus  0 & \pminus  0 & \pminus  0 & \pminus  0 & \pminus  0 & \pminus  0 & \pminus  0 & \pminus  0 \\ \pminus
 0 & \pminus  0 & \pminus  0 & \pminus  0 & \pminus  0 & \pminus  0 & \pminus  0 & \pminus  0 & \pminus  0 & \pminus  0 & \pminus  0 & \pminus  0 & \pminus  0 & \pminus  0 & \pminus  0 & \pminus  0 & \pminus  0 & \pminus  0 & \pminus  0 & \pminus  0 & \pminus  0 & \pminus  0 & \pminus  0 & \pminus  0 & \pminus  0 & \pminus  0 & \pminus  0 & \pminus  0 & \pminus  0 & \pminus  0 & \pminus  0 & \pminus  0 & \pminus  0 & \pminus  0 & \pminus  0 & \pminus  0 & \pminus  0 & \pminus  0 & \pminus  0 & \pminus  0 \\
\end{array}
\right)
  },
\end{align}
\begin{align}
  A_{8} &=\scalemath{0.5}{
  \left(
\begin{array}{cccccccccccccccccccccccccccccccccccccccc}
\pminus  0 & \pminus  0 & \pminus  0 & \pminus  0 & \pminus  0 & \pminus  0 & \pminus  0 & \pminus  0 & \pminus  0 & \pminus  0 & \pminus  0 & \pminus  0 & \pminus  0 & \pminus  0 & \pminus  0 & \pminus  0 & \pminus  0 & \pminus  0 & \pminus  0 & \pminus  0 & \pminus  0 & \pminus  0 & \pminus  0 & \pminus  0 & \pminus  0 & \pminus  0 & \pminus  0 & \pminus  0 & \pminus  0 & \pminus  0 & \pminus  0 & \pminus  0 & \pminus  0 & \pminus  0 & \pminus  0 & \pminus  0 & \pminus  0 & \pminus  0 & \pminus  0 & \pminus  0 \\ \pminus
 0 & \minus 4 & \pminus  0 & \pminus  0 & \pminus  0 & \pminus  0 & \pminus  0 & \pminus  0 & \pminus  0 & \pminus  0 & \pminus  0 & \pminus  0 & \pminus  0 & \pminus  0 & \pminus  0 & \pminus  0 & \pminus  0 & \pminus  0 & \pminus  0 & \pminus  0 & \pminus  0 & \pminus  0 & \pminus  0 & \pminus  0 & \pminus  0 & \pminus  0 & \pminus  0 & \pminus  0 & \pminus  0 & \pminus  0 & \pminus  0 & \pminus  0 & \pminus  0 & \pminus  0 & \pminus  0 & \pminus  0 & \pminus  0 & \pminus  0 & \pminus  0 & \pminus  0 \\ \pminus
 0 & \pminus  0 & \minus 2 & \pminus  0 & \pminus  0 & \pminus  0 & \pminus  0 & \pminus  0 & \pminus  0 & \pminus  0 & \pminus  0 & \pminus  0 & \pminus  0 & \pminus  0 & \pminus  0 & \pminus  0 & \pminus  0 & \pminus  0 & \pminus  0 & \pminus  0 & \pminus  0 & \pminus  0 & \pminus  0 & \pminus  0 & \pminus  0 & \pminus  0 & \pminus  0 & \pminus  0 & \pminus  0 & \pminus  0 & \pminus  0 & \pminus  0 & \pminus  0 & \pminus  0 & \pminus  0 & \pminus  0 & \pminus  0 & \pminus  0 & \pminus  0 & \pminus  0 \\
\end{array}
\right)
  },
 \end{align}
\begin{align}
  A_{9} &=\scalemath{0.5}{
  \left(
\begin{array}{cccccccccccccccccccccccccccccccccccccccc}
\pminus 0 & \pminus 0 & \pminus 0 & \pminus 0 & \pminus 0 & \pminus 0 & \pminus 0 & \pminus 0 & \pminus 0 & \pminus 0 & \pminus 0 & \pminus 0 & \pminus 0 & \pminus 0 & \pminus 0 & \pminus 0 & \pminus 0 & \pminus 0 & \pminus 0 & \pminus 0 & \pminus 0 & \pminus 0 & \pminus 0 & \pminus 0 & \pminus 0 & \pminus 0 & \pminus 0 & \pminus 0 & \pminus 0 & \pminus 0 & \pminus 0 & \pminus 0 & \pminus 0 & \pminus 0 & \pminus 0 & \pminus 0 & \pminus 0 & \pminus 0 & \pminus 0 & \pminus 0 \\ \pminus
 0 & \pminus 0 & \pminus 0 & \pminus 0 & \pminus 0 & \pminus 0 & \pminus 0 & \pminus 0 & \pminus 0 & \pminus 0 & \pminus 0 & \pminus 0 & \pminus 0 & \pminus 0 & \pminus 0 & \pminus 0 & \pminus 0 & \pminus 0 & \pminus 0 & \pminus 0 & \pminus 0 & \pminus 0 & \pminus 0 & \pminus 0 & \pminus 0 & \pminus 0 & \pminus 0 & \pminus 0 & \pminus 0 & \pminus 0 & \pminus 0 & \pminus 0 & \pminus 0 & \pminus 0 & \pminus 0 & \pminus 0 & \pminus 0 & \pminus 0 & \pminus 0 & \pminus 0 \\ \pminus
 0 & \pminus 0 & \pminus 0 & \pminus 0 & \pminus 0 & \pminus 0 & \pminus 0 & \pminus 0 & \pminus 0 & \pminus 0 & \pminus 0 & \pminus 0 & \pminus 0 & \pminus 0 & \pminus 0 & \pminus 0 & \pminus 0 & \pminus 0 & \pminus 0 & \pminus 0 & \pminus 0 & \pminus 0 & \pminus 0 & \pminus 0 & \pminus 0 & \pminus 0 & \pminus 0 & \pminus 0 & \pminus 0 & \pminus 0 & \pminus 0 & \pminus 0 & \pminus 0 & \pminus 0 & \pminus 0 & \pminus 0 & \pminus 0 & \pminus 0 & \pminus 0 & \pminus 0 \\ \pminus
 0 & \pminus 0 & \pminus 0 & \pminus 0 & \pminus 0 & \pminus 0 & \pminus 0 & \pminus 0 & \pminus 0 & \pminus 0 & \pminus 0 & \pminus 0 & \pminus 0 & \pminus 0 & \pminus 0 & \pminus 0 & \pminus 0 & \pminus 0 & \pminus 0 & \pminus 0 & \pminus 0 & \pminus 0 & \pminus 0 & \pminus 0 & \pminus 0 & \pminus 0 & \pminus 0 & \pminus 0 & \pminus 0 & \pminus 0 & \pminus 0 & \pminus 0 & \pminus 0 & \pminus 0 & \pminus 0 & \pminus 0 & \pminus 0 & \pminus 0 & \pminus 0 & \pminus 0 \\
\end{array}
\right)
  },
 \end{align}
\begin{align}
~~~~~~~~A_{10} &=\scalemath{0.50}{
  \left(
\begin{array}{cccccccccccccccccccccccccccccccccccccccc}
\pminus 0 & \pminus 0 & \pminus 0 & \pminus 0 & \pminus 0 & \pminus 0 & \pminus 0 & \pminus 0 & \pminus 0 & \pminus 0 & \pminus 0 & \pminus 0 & \pminus 0 & \pminus 0 & \pminus 0 & \pminus 0 & \pminus 0 & \pminus 0 & \pminus 0 & \pminus 0 & \pminus 0 & \pminus 0 & \pminus 0 & \pminus 0 & \pminus 0 & \pminus 0 & \pminus 0 & \pminus 0 & \pminus 0 & \pminus 0 & \pminus 0 & \pminus 0 & \pminus 0 & \pminus 0 & \pminus 0 & \pminus 0 & \pminus 0 & \pminus 0 & \pminus 0 & \pminus 0 \\ \pminus
 0 & \pminus 0 & \pminus 0 & \pminus 0 & \pminus 0 & \pminus 0 & \pminus 0 & \pminus 0 & \pminus 0 & \pminus 0 & \pminus 0 & \pminus 0 & \pminus 0 & \pminus 0 & \pminus 0 & \pminus 0 & \pminus 0 & \pminus 0 & \pminus 0 & \pminus 0 & \pminus 0 & \pminus 0 & \pminus 0 & \pminus 0 & \pminus 0 & \pminus 0 & \pminus 0 & \pminus 0 & \pminus 0 & \pminus 0 & \pminus 0 & \pminus 0 & \pminus 0 & \pminus 0 & \pminus 0 & \pminus 0 & \pminus 0 & \pminus 0 & \pminus 0 & \pminus 0 \\ \pminus
 0 & \pminus 0 & \pminus 0 & \pminus 0 & \pminus 0 & \pminus 0 & \pminus 0 & \pminus 0 & \pminus 0 & \pminus 0 & \pminus 0 & \pminus 0 & \pminus 0 & \pminus 0 & \pminus 0 & \pminus 0 & \pminus 0 & \pminus 0 & \pminus 0 & \pminus 0 & \pminus 0 & \pminus 0 & \pminus 0 & \pminus 0 & \pminus 0 & \pminus 0 & \pminus 0 & \pminus 0 & \pminus 0 & \pminus 0 & \pminus 0 & \pminus 0 & \pminus 0 & \pminus 0 & \pminus 0 & \pminus 0 & \pminus 0 & \pminus 0 & \pminus 0 & \pminus 0 \\ \pminus
 0 & \pminus 0 & \pminus 0 & \pminus 0 & \pminus 0 & \pminus 0 & \pminus 0 & \pminus 0 & \pminus 0 & \pminus 0 & \pminus 0 & \pminus 0 & \pminus 0 & \pminus 0 & \pminus 0 & \pminus 0 & \pminus 0 & \pminus 0 & \pminus 0 & \pminus 0 & \pminus 0 & \pminus 0 & \pminus 0 & \pminus 0 & \pminus 0 & \pminus 0 & \pminus 0 & \pminus 0 & \pminus 0 & \pminus 0 & \pminus 0 & \pminus 0 & \pminus 0 & \pminus 0 & \pminus 0 & \pminus 0 & \pminus 0 & \pminus 0 & \pminus 0 & \pminus 0 \\
\end{array}
\right)
  }.
 \end{align}

Both matrices $A_{11}$ and $A_{12}$ have only two non vanishing entries, with
\bea
(A_{11})_{38,35}&=&(A_{12})_{40,36}=-12,\nonumber\\
(A_{11})_{38,38}&=&(A_{12})_{40,40}=-4.\nonumber
\eea

\vspace{3ex}
\bibliographystyle{JHEP}
\bibliography{references}

\end{document}